\documentstyle[epsf,aps,preprint]{revtex}

\begin{document}
\title{Two-stage superconducting-quantum-interference-device
       amplifier in a high-Q gravitational wave transducer}
\author{Gregory~M.~Harry$^{a)}$\footnotetext{$^{a)}$Present address: Department
        of Physics, Syracuse University, Syracuse, New York
        13244-1130}, Insik~Jin, Ho~Jung~Paik,
        Thomas~R.~Stevenson$^{b)}$
        \footnotetext{$^{b)}$Present address: Department of Applied Physics,
        Yale University, PO Box 208284, New Haven, Connecticut
        06520-8284},  and Frederick~C.~Wellstood}
\address{Center for Superconductivity Research, Department of
         Physics, University of Maryland, College~Park, Maryland
         20742}
\date{\today}
\maketitle

\begin{abstract}
We report on the total noise from an inductive motion transducer
for a gravitational-wave antenna.  The transducer uses a
two-stage SQUID amplifier and has a noise temperature of 1.1~mK,
of which 0.70~mK is due to back-action noise from the
superconducting quantum interference device (SQUID) chip.  The
total noise includes thermal noise from the transducer mass,
which has a measured Q of $2.60 \times 10^{6}$. The noise
temperature exceeds the expected value of 3.5~$\mu$K by a factor
of 200, primarily due to voltage noise at the input of the
SQUID.  Noise from flux trapped on the chip is found to be the
most likely cause.
\end{abstract}

\pacs{PACS number(s): 85.25.Dq, 74.40.+k, 04.80.Nn}

Detection of gravitational waves from astronomical sources requires
extremely low noise antennas and amplifiers~\cite{lsu}.  The dominant noise
source is the first stage electrical amplifier, which has typically been
made from a superconducting quantum interference device (SQUID).
Our previous work on gravitational wave transducers using
a commercial SQUID from Quantum Design of San Diego California found a
noise temperature of 3.9~mK, with 1.2~mK attributed to SQUID back-action
noise~\cite{thomasamaldi}.  Here we report noise measurements for an
integrated two-SQUID system in which one SQUID amplifies the output of
another.  We use the interaction between the SQUID input and the high-Q
transducer circuit to distinguish different noise sources internal to
the SQUIDs.

The first-stage SQUID is used as an electrical amplifier and immediately
follows the transducer, which is connected to the gravitational-wave antenna.
The transducer used for these measurements was a Paik-style~\cite{paik}, inductive,
resonant mass.  Figure~\ref{fig:antenna}(a) shows a schematic of
the antenna with a transducer mass.
There are two coils of superconducting wire on either side of the
transducer's proof mass, as shown in Fig.~\ref{fig:antenna}(b). Conservation of
magnetic flux in a superconducting circuit requires that persistent
current stored in these coils changes with inductance.  This signal
current is shunted to the transformer with primary $L_{t1}$.  The secondary, $L_{t2}$,
of this transformer is connected to the input of the SQUID chip.


The transducer was made from a round plate of niobium out of which
circular
grooves were milled on both faces, defining a central mass.  The remaining thin
niobium annulus acts as the mechanical spring connecting the
central proof mass to the case.  The case would be bolted rigidly to the
antenna during gravitational-wave searches.  The proof mass was electropolished
in an acid solution and
was heat treated to 1500$^\circ$~C to improve the quality factor of the
resonance.  Two other niobium plates were then bolted
to either side of the proof mass and contain the sensing coils $L_{1}$ and $L_{2}$.
Measurements revealed 10~$\mu$m gaps between the
coils and the proof mass at room temperature and 25~$\mu$m gaps at the
operating temperature of 4.2~K.

The SQUID amplifier was comprised of two SQUIDs, the first
serving as a preamplifier for the second~\cite{insik}.  Both SQUIDs had
junctions made from Nb-Al/AlO$_{x}$-Nb trilayers and were impedance
matched to the transducer circuit using a 40:1, thin-film, on-chip
transformer.  The first SQUID was
kept in a flux-locked loop by modulating the second SQUID with a 500~kHz
square-wave signal.  The demodulated signal was negatively
fed back to the first SQUID to linearize the amplifier response.
A second feedback loop was employed to keep the flux gain between the
SQUIDs at a maximum~\cite{GR14}  (see Figure~\ref{feedback}). The
additional loop modulated the second SQUID with a small amplitude flux
signal at 8~kHz.  This signal was demodulated at 16~kHz and the resulting
low frequency signal was negatively fed back to the second SQUID.  Using
the second harmonic of the input signal as the source of the feedback flux
made this loop sensitive to the second
derivative of the inverse of the function
\begin{equation}
\Phi_{2} ( \Phi_{1} ) = \frac{G_{\Phi}}{2 \pi} \sin \{ \frac{2
\pi}{\Phi_{0}} [ \Phi_{1} + \Phi_{\mathrm{B1}}(t)]\} +
\Phi_{\mathrm{B2}}(t),
\end{equation}
where $\Phi_{2}$ is the flux in the second SQUID, $\Phi_{1}$ is
the flux in the first SQUID, $G_{\Phi}$ is the maximal flux gain
between the two SQUIDs, $\Phi_{0}$ is the quantum of magnetic
flux, and $\Phi_{\mathrm{B1}}(t)$ and $\Phi_{\mathrm{B2}}(t)$ are
the (possibly time-varying) background fluxes in the first and
second SQUID, respectively. This process is equivalent to
maximizing the flux gain $d \Phi_{2}/d \Phi_{1}$ by shifting the
background fluxes.  Determining and then maintaining this
maximized flux gain against changing an external background flux
$\Phi_{\mathrm{B2}}(t)$ is necessary to minimize the effect of
noise from both SQUIDs.


To make noise measurements, the SQUID amplifier was connected to
the transducer (the transducer was not attached to an antenna in
these experiments but was instead suspended from a vibration
isolator). A current of 6.5025~A was stored in the  two coils
surrounding the proof mass. The noise was recorded as voltage
across the feedback resistor $R_{\mathrm{fb}}$.
Figure~\ref{fig:noise} shows the resulting root-mean-square flux
noise spectrum in a 62.5~Hz bandwidth around the transducer
resonance at 892~Hz.  We use the observed contrast between
constructive and destructive interference between correlated
noise above and below the transducer resonance frequency, evident
in Fig.~\ref{fig:noise}, to make qualitative and quantitative
inferences about the magnitude and source of back-action noise.


Two sources of noise were expected to dominate the total noise:
additive noise from the SQUID amplifier and thermal force noise
from dissipation in the proof mass.  The expected signal-to-noise
ratio density, $r(\omega)$, can be written~\cite{Price},
\begin{equation}
r(\omega) = \frac{2 \Re[Z_{n}] \mu \omega^{2}}{k_{B} T_{n} |k -
\mu \omega^{2} + j \omega Z_{n}|^{2}},
\label{snrdensity}
\end{equation}
where $Z_{n}$ is the mechanical noise impedance, $\mu$ is the
reduced mass of the proof mass and case mass, $k$ is the spring
constant between the proof mass and the case, $k_{B}$ is
Boltzmann's constant, and $T_{n}$ is the noise temperature. This
quantity $r(\omega)$ represents the sensitivity of the transducer
per unit energy deposited in the transducer resonance by a
signal.  A more complete discussion of its derivation and use can
be found in~\cite{Price} and~\cite{thesis}.

After calibrating the transducer response to a mechanical signal,
we used Eq.~\ref{snrdensity} to describe the total noise in terms
of a noise temperature $T_{n}$ and a complex noise impedance
$Z_{n}$. The values of $T_{n}$ and $Z_{n}$ describe the total
force and velocity noises and their correlation and can be found
from fitting the noise data to Eq.~\ref{snrdensity}.  We found
from this fit
\begin{eqnarray}
T_{n} & = & 1.08 \times 10^{-3}~\mathrm{K}, \\
Z_{n} & = & (16.9 + 4.24~j)~\mathrm{kg/s}.
\end{eqnarray}
The double-sided spectral density of force noise, velocity noise,
and their correlation can then be obtained from these
parameters~\cite{Price}.  Through additional calibration
measurements~\cite{thesis}, we were able to characterize the
electromechanical circuit parameters in the transducer and SQUID
chip to allow us to quantitatively relate $T_{n}$ and $Z_{n}$ to
various possible mechanical and electrical noise sources internal
to the transducer and SQUIDs.

The back-action noise due to the SQUID was determined by
subtracting the thermal force noise from the total force noise
observed.  The thermal noise arises from the dissipation of the
transducer resonance.  The magnitude of the thermal force noise
can be predicted using the fluctuation-dissipation theorem and
the measured exponential decay time of the resonance.  The Q for
this mode is measured directly from the damped oscillation and
was found to be
\begin{equation}
Q =  2.60 \times 10^{6},
\end{equation}
which includes contributions from the mechanical spring and the
electrical spring created by the stored currents.  This Q-value has been corrected for
cold damping produced by the SQUID feedback loop~\cite{thomasamaldi} so that it gives the
passive dissipation in thermal equilibrium at the transducer
temperature.  By studying the dependence of Q on stored current, the
total Q could be broken down into mechanical and electrical components~\cite{paik}:
\begin{eqnarray}
Q_{m} & = & 3.15 \times 10^{6}, \\
Q_{e} & = & 2.52 \times 10^{5}.
\end{eqnarray}
The thermal force noise can be found using the total $Q$ from
\begin{equation}
S_{f} = 2 k_{B} T \frac{\omega_{0} \mu}{Q},
\end{equation}
where $T$ is the measured physical temperature,  $\omega_{0}$ is the resonance frequency,
$\mu$ is the reduced mass of the proof mass and the case, and $Q$
is the measured, passive Q of the resonance.  After subtracting this thermal noise,
the noise temperature of the SQUID amplifier was found to be
\begin{equation}
T_{s}  =  6.99 \times 10^{-4}~\mathrm{K}.
\end{equation}
Using a circuit model~\cite{thomasamaldi,thesis}
for the transducer and SQUID input circuit, we calculated the SQUID's
electrical noise impedance:
\begin{equation}
Z_{s}  =  (5.9 \times 10^{-3} - 8.9 \times 10^{-6}~j)~\Omega.
\end{equation}


We also performed tests on the SQUID without the transducer attached.  The
input port was left open-circuited and the entire SQUID chip was contained
in a niobium box.  The SQUID's energy sensitivity in this configuration was
$9.22 \times 10^{-6}$~K.  We note that with only one port on the
amplifier available, the noise can only be expressed as a single number
and a true noise temperature can not be calculated.

We used the method of Clarke, Tesche, and Giffard~\cite{CTG}, as extended by
Martinis and Clarke~\cite{martinis}, to calculate the
minimum or true noise temperature of the SQUID expected from the Johnson
noise of its shunt resistors:
\begin{equation}
T_{\mathrm{CTG}} = 3.5 \times 10^{-6}~\mathrm{K}.
\end{equation}
Table~\ref{noise} presents the Clarke-Tesche-Giffard (CTG) predictions for each noise
component as well as the experimental results.  The data
is presented as both the experimental result and as a minimum limit derived
from the data. These limits come from using the observed high/low frequency asymmetry
seen in the noise spectrum while assuming the thermal noise is large
enough that an accurate force noise subtraction cannot be done.
The voltage noise and current noise must be at least as large as this limiting
case.


The measured noise temperature is a factor of 200 above the
expected noise of $T_{\mathrm{CTG}}$, thereby reducing the
sensitivity of the transducer through Eq.~\ref{snrdensity}. This
noise is almost exclusively due to voltage noise at the input of
the SQUID as the excess noise is observable only when the SQUID is
coupled to the high-Q transducer. We examined possible
explanations for this excess voltage noise. Flux creep in the
large sensing coils $L_{1}$ and $L_{2}$ (see
Figure~\ref{fig:antenna}~(b)) was eliminated as a possible source
based on the predicted signature of the two noise components.
Noise from a varying magnitude RF signal was also
considered~\cite{thomasamaldi}  but was rejected because we
expect an accompanying large current noise, which was not
observed.  Noise from flux lines moving between pinning sites in
the on-chip transformer $L_{s4}$ was modeled using the method of
Ferrari~{\em et al.}~\cite{ferrari}.  We found that the product
\begin{equation}
n S_{r} = 9.1 \times 10^{-9}~\mathrm{Hz}^{-1},
\end{equation}
where $n$ is the flux vortex density and $S_{r}$ is the spectral
density of the flux's radial motion, would give rise to the observed
voltage noise.  This same $n S_{r}$ would predict values of
\begin{eqnarray}
 S_{I} (\omega_{0} L)/k_{B} & = & 1.8 \pm 0.7~\mu\mathrm{K}, \\
|S_{VI}|/k_{B} & = & 80 \pm 30~\mu\mathrm{K}.
\end{eqnarray}
The predicted value of the current noise is below the CTG value, so on-chip
flux noise would not be significant.  The correlation noise agrees
within the error bars with the experimental limit in Table~\ref{noise}.

While the back-action noise we observe would allow only modest
sensitivity improvement for detectors operating at 4
K~\cite{lsu}, the potential for millikelvin
detectors~\cite{auriga,nautilus} may be better. We were unable to
measure the SQUID in a transducer at millikelvin temperatures,
but separate noise measurements on a similar chip in a dilution
refrigerator~\cite{IEEE,Jin97} indicate flux motion noise may be
much less at 90~mK, depending on the source of the flux.  Our
SQUID is being considered for use in a multimode transducer on
the ALLEGRO detector~\cite{lsu} and, with further research, it
may be possible to improve the 4 K performance by using a heater
to expel flux.

This work was supported by grant PHY93-12229 from the National Science Foundation
and by the Gravitational Wave Lab at Louisiana State University.

\begin{table}[p]
\caption{Voltage and current noise at the input of the first
SQUID.  The three columns represent the experimentally determined
values, the minimum experimental limits, and the values predicted
from Clarke-Tesche-Giffard theory.  All noise values are
presented in temperature units using the inductance of the pickup
coils as expressed through the two transformers, $L=740$~pH.}
\label{noise}
\begin{center}
\begin{tabular}{rrrr}
Parameter & Experimental Value & Limit & CTG Value \\
\hline
$S_{V}/\omega_{0} L k_{B}$ & 3.3~mK & 400~$\mu$K & 0.8~$\mu$K \\
$S_{I} (\omega_{0} L/k_{B})$ & 490~$\mu$K & 20~$\mu$K & 50~$\mu$K \\
$-\Im[S_{VI}]/k_{B}$        & 1.1~mK & 100~$\mu$K & 5.3~$\mu$K
\end{tabular}
\end{center}
\end{table}

\begin{figure}[p]
\caption{(a) Mechanical model of antenna with transducer
mass.  The equivalent mass of the antenna, $M_{1}$, is connected by a
spring with constant $k_{1}=M_{1} \omega_{0}^{2}$ to mechanical ground and
the
antenna is connected to the proof mass, $M_{2}$, by a spring with constant
$k_{2}= M_{2} \omega_{0}^{2}$. The force from the gravitational wave, $F$, acts
between mechanical ground and the antenna mass.
(b) Circuit diagram of the transducer.  The
proof mass can move between the
two pickup coils, $L_{1}$ and $L_{2}$.  The signal current from this
motion passes through two transformers to the first SQUID.  All of the
wiring in the transducer circuit is superconducting.}
\label{fig:antenna}
\end{figure}

\begin{figure}[p]
\caption{Dual feedback loops used to control the two-SQUID system.  The
flux-locking loop is at 500~kHz.  The modulation is sent to the second
SQUID and the low frequency feedback is sent to the first SQUID.  The
loop to maximize the flux gain modulates the second SQUID at 8~kHz, but
is demodulated at 16~kHz.}
\label{feedback}
\end{figure}

\begin{figure}[p]
\caption{Noise spectral density near the transducer resonance expressed as
flux in the first SQUID.  The asymmetry in the peak indicates an excess
of back-action noise.}
\label{fig:noise}
\end{figure}

\pagebreak
Figure 1 - G.~M.~Harry {\em et al.}

\hspace{-1.0em}(a)
\begin{center}
\epsfxsize=6in
\leavevmode \epsfbox{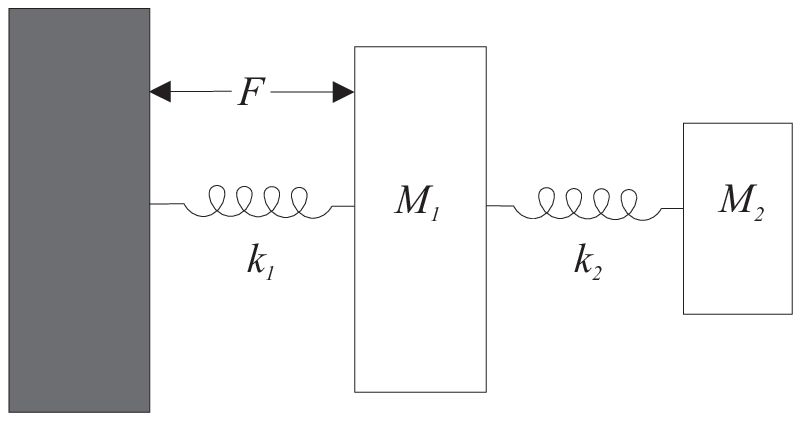}
\end{center}
(b)
\begin{center}
\epsfxsize=6in
\leavevmode \epsfbox{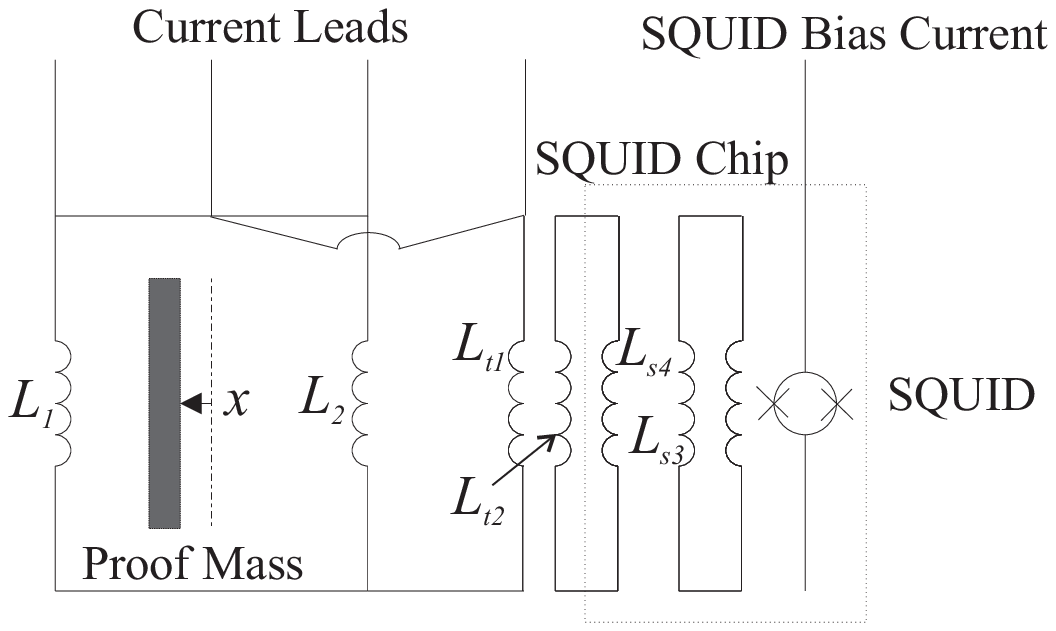}
\end{center}

\pagebreak
Figure 2 - G.~M.~Harry {\em et al.}
\begin{center}
\epsfxsize=6in \leavevmode \epsfbox{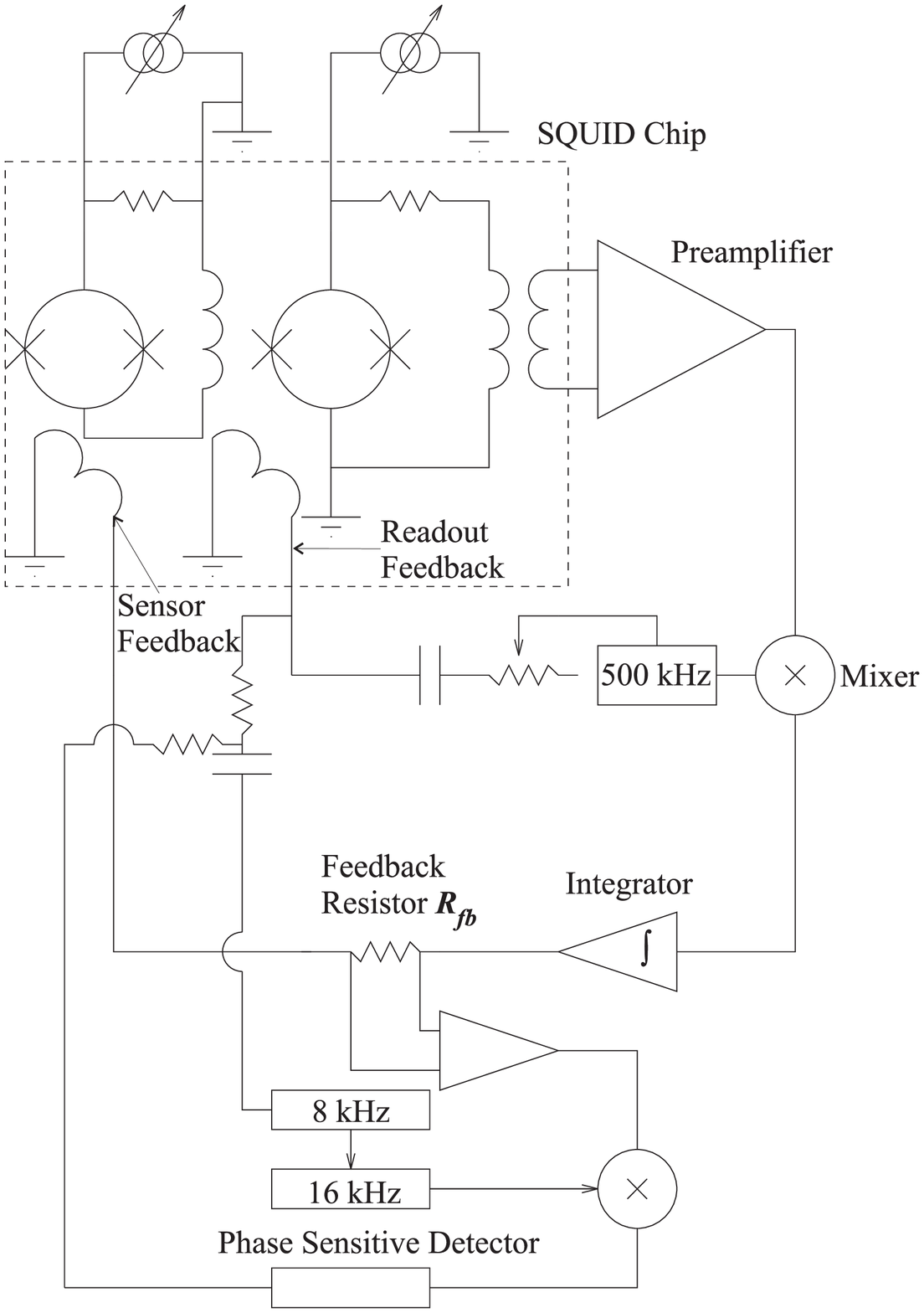}
\end{center}

\pagebreak
Figure 3 - G.~M.~Harry {\em et al.}
\begin{center}
\epsfxsize=6in \leavevmode \epsfbox{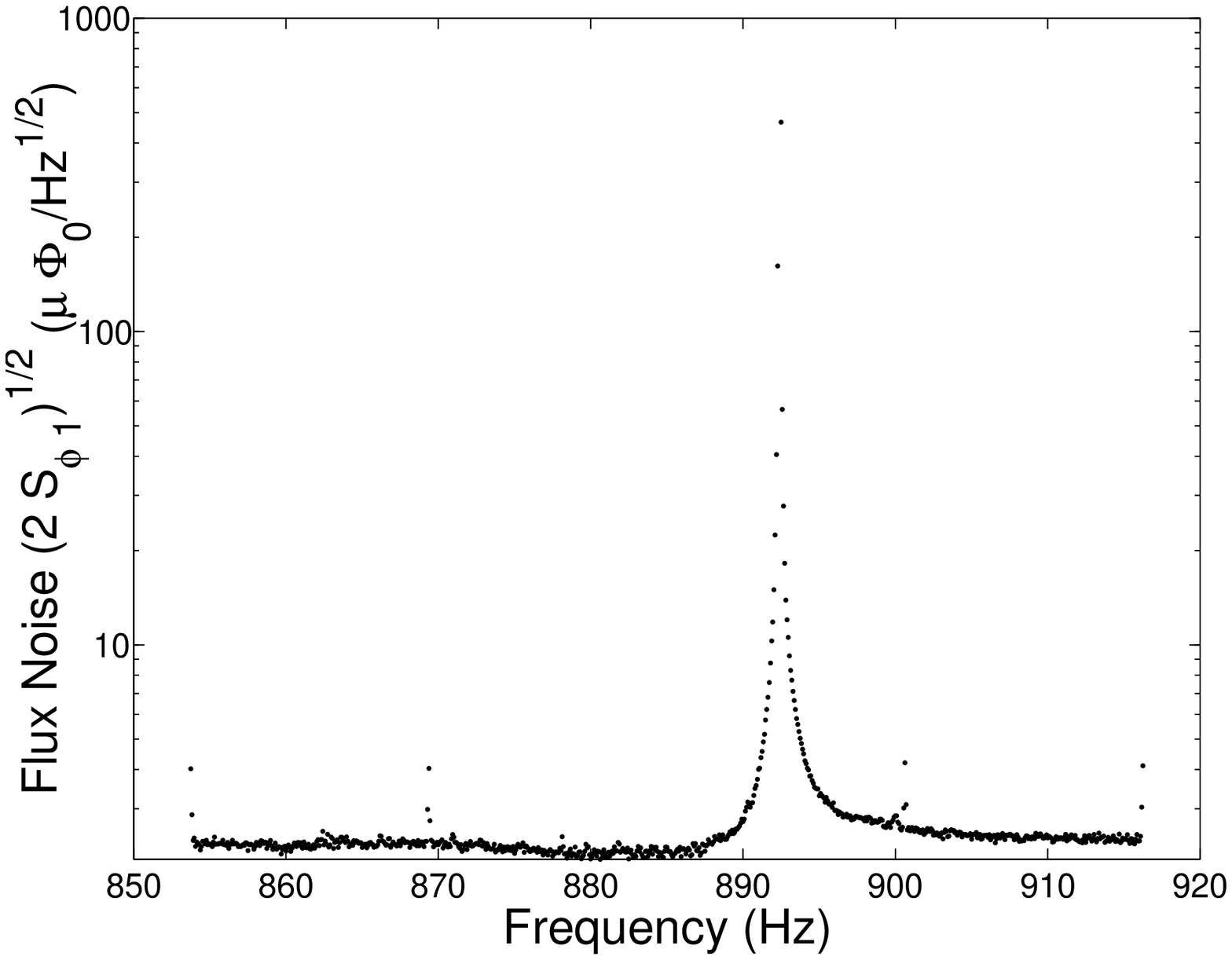}
\end{center}

\end{document}